\documentclass[10pt,conference,letterpaper]{IEEEtran}

\usepackage{graphicx}
\usepackage{balance}  % for  \balance command ON LAST PAGE  (only there!)
\usepackage{multirow,tabularx, array, color, url, subfig}
\usepackage{algorithm}
\usepackage{algpseudocode}
\usepackage{hyperref}
\usepackage{tabularx}
\usepackage[utf8]{inputenc}
\usepackage{amsfonts}
\usepackage{amsmath}
\usepackage{threeparttable}
\usepackage{listings}
\usepackage{color}
\usepackage{lipsum}

\usepackage{enumitem}

\setlist{itemsep=0pt}

\definecolor{codegreen}{rgb}{0,0.6,0}
\definecolor{codegray}{rgb}{0.5,0.5,0.5}
\definecolor{codepurple}{rgb}{0.58,0,0.82}
\definecolor{backcolour}{rgb}{0.95,0.95,0.92}
\lstdefinestyle{mystyle}{
    backgroundcolor=\color{backcolour},   
    commentstyle=\color{codegreen},
    keywordstyle=\color{magenta},
    numberstyle=\tiny\color{codegray},
    stringstyle=\color{codepurple},
    basicstyle=\footnotesize,
    breakatwhitespace=false,         
    breaklines=true,                 
    captionpos=b,                    
    keepspaces=true,                 
    numbers=left,                    
    numbersep=5pt,                  
    showspaces=false,                
    showstringspaces=false,
    showtabs=false,                  
    tabsize=2
}

\newcommand\HH{% horrible hack
  \global\let\savedtextbullet\textbullet
  \gdef\textbullet{%
    \par\noindent\savedtextbullet\global\let\textbullet\savedtextbullet
  }%
}

\begin{document}

\title{Matrix Factorization on GPUs with Memory Optimization and Approximate Computing}
\author{%
% author names are typeset in 11pt, which is the default size in the author block
{Wei Tan{\small $~^{\#}$}, Shiyu Chang{\small $~^{*}$}, Liana Fong{\small $~^{*}$}, Cheng Li{\small $~^{**}$}, Zijun Wang{\small $~^{*}$}, Liangliang Cao{\small $~^{\#\#}$}
}
% add some space between author names and affils
\vspace{1.6mm}\\
\fontsize{10}{10}\selectfont\itshape
$^{\#}$\,Citadel (Work done while author was with IBM.)\\
\fontsize{9}{9}\selectfont\ttfamily\upshape
weitan@ieee.org
% add some space between email and affil
\vspace{1.2mm}\\
\fontsize{10}{10}\selectfont\rmfamily\itshape
$^{*}$\,IBM T. J. Watson Research Center\\
\fontsize{9}{9}\selectfont\ttfamily\upshape
shiyu.chang@ibm.com, llfong@us.ibm.com, zijun.wang@ibm.com
\vspace{1.2mm}\\
\fontsize{10}{10}\selectfont\rmfamily\itshape
$^{**}$\,University of Illinois at Urbana-Champaign
\\
\fontsize{9}{9}\selectfont\ttfamily\upshape
cli99@illinois.edu
\vspace{1.2mm}\\
\fontsize{10}{10}\selectfont\rmfamily\itshape
$^{\#\#}$\,HelloVera.AI
\\
\fontsize{9}{9}\selectfont\ttfamily\upshape
llc@hellovera.ai 
}
\maketitle

% declaration of the new block
\algblock{ParFor}{EndParFor}
% customising the new block
\algnewcommand\algorithmicparfor{\textbf{parfor}}
\algnewcommand\algorithmicpardo{\textbf{do}}
\algnewcommand\algorithmicendparfor{\textbf{end\ parfor}}
\algrenewtext{ParFor}[1]{\algorithmicparfor\ #1\ \algorithmicpardo}
\algrenewtext{EndParFor}{\algorithmicendparfor}
\newcommand{\algname}{\textsc{cuMFals }}
\newcommand{\algnamens}{\textsc{cuMFals}}

\begin{abstract}
Matrix factorization (MF) discovers latent features from observations, which has shown great promises in the fields of collaborative filtering, data compression, feature extraction, word embedding, \emph{etc.} While many problem-specific optimization techniques have been proposed, alternating least square (ALS) remains popular due to its general applicability (\emph{e.g.} easy to handle positive-unlabeled inputs), 
fast convergence and parallelization capability. Current MF implementations are either optimized for a single machine or with a need of a large computer cluster but still are insufficient. This is because a single machine provides limited compute power for large-scale data while multiple machines suffer from the network communication bottleneck. 

To address the aforementioned challenge, accelerating ALS on graphics processing units (GPUs) is a promising direction. We propose the novel approach in enhancing the MF efficiency via both \textbf{memory optimization} and \textbf{approximate computing}. The former exploits GPU memory hierarchy to increase data reuse, while the later reduces unnecessary computing without hurting the convergence of learning algorithms. Extensive experiments on large-scale datasets show that our solution not only outperforms the competing CPU solutions by a large margin 
but also has a \textbf{2x-4x} performance gain compared to 
the state-of-the-art GPU solutions. 
Our implementations are open-sourced and publicly available. 
\end{abstract}

\section{Introduction}
\label{sec:intro}
% \begin{figure}
% \center{\includegraphics[width=\linewidth]
%         {figs/mf.pdf}}
%  \caption{Matrix factorization.}
% \label{fig:mf}
% \end{figure}
% Matrix factorization (MF) factorizes a matrix $R \in \mathbb R^{m\times n}$ (with $N_z$ non-zero elements) into two dense matrices $X \in \mathbb R^{m\times f}$ and $\Theta \in \mathbb R^{n\times f}$, such that $R\approx X \cdot \Theta^{T}$. For $1\leq u\leq m$ and $1\leq v\leq n$ and $r_{uv}$ is the $(i,j)$ entry of $R$, $r_{uv}\approx \boldsymbol{x}_u^T \cdot \boldsymbol\theta_v$, where $f$ is the given dimension of factors, and $\boldsymbol{x}_u, \boldsymbol{\theta}_v\in \mathbb R^f$ are the $u^{th}$ column of $X^T$ and the $v^{th}$ column of $\Theta^T$, respectively.
%That is, the inner product of the two entity features $\boldsymbol{x}_u$ and $\boldsymbol\theta_v$ determines the interaction degree $r_{uv}$ between them.

Matrix factorization (MF) is one of the most important data mining techniques due to its implementation simplicity and broad applicability. For instance, MF is the core of modern recommender systems~\cite{mf-computer09,spotify,facebook15}. MF has also been widely used in compressing large models (\emph{e.g.} deep neural networks) for mobile usage~\cite{NIPS2014_5484}, calculating word embedding \cite{mf-computer09,pennington2014glove}, \emph{etc.}  However, the big data processing, with massive data is generated at an unprecedented rate, demands further acceleration of MF.  For example, the number of active users of Facebook exceed 1.860 billion in the fourth quarter of 2016\footnote{\url{https://ibm.biz/BdstmU}}. Solving MF efficiently under such a large scale challenges many existing solutions. 

Although many studies~\cite{factorbird14,libmf-13,libmf++,sparkler13,nomad14,DSGD-kdd11,ccd++-icdm12,hogwild-nips11,kdd15mf} have been conducted to accelerate MF, they are still insufficient to process large scale data set. These methods either use multiple threads on one machine or multiple processes on distributed systems. The former one uses shared memory which is efficient but hard to handle big data in real-world settings.  On the other hand, the communication cost becomes the major bottleneck in distributed systems, which significantly reduces its efficiency in terms of aggregated floating point operations per second (FLOPS).  Nevertheless, with recent successes of deep learning \cite{costshpc2013} using graphics processing units (GPUs), there comes a new venue for expediting other data mining algorithms \cite{bidmach2015,hpdc2016}.  GPU has superior compute power and memory bandwidth compared to CPU~\cite{hennessy2011computer}. Moreover, GPUs on one server can leverage interconnections such as NVLink~\cite{nvlink} (40 GB/s per link with four links per GPU) which is much faster than any existing network. Therefore, we consider to solve the problem of MF using the alternating least square (ALS) method on GPUs.

\begin{figure}[t]
\center{\includegraphics[width=\linewidth]
        {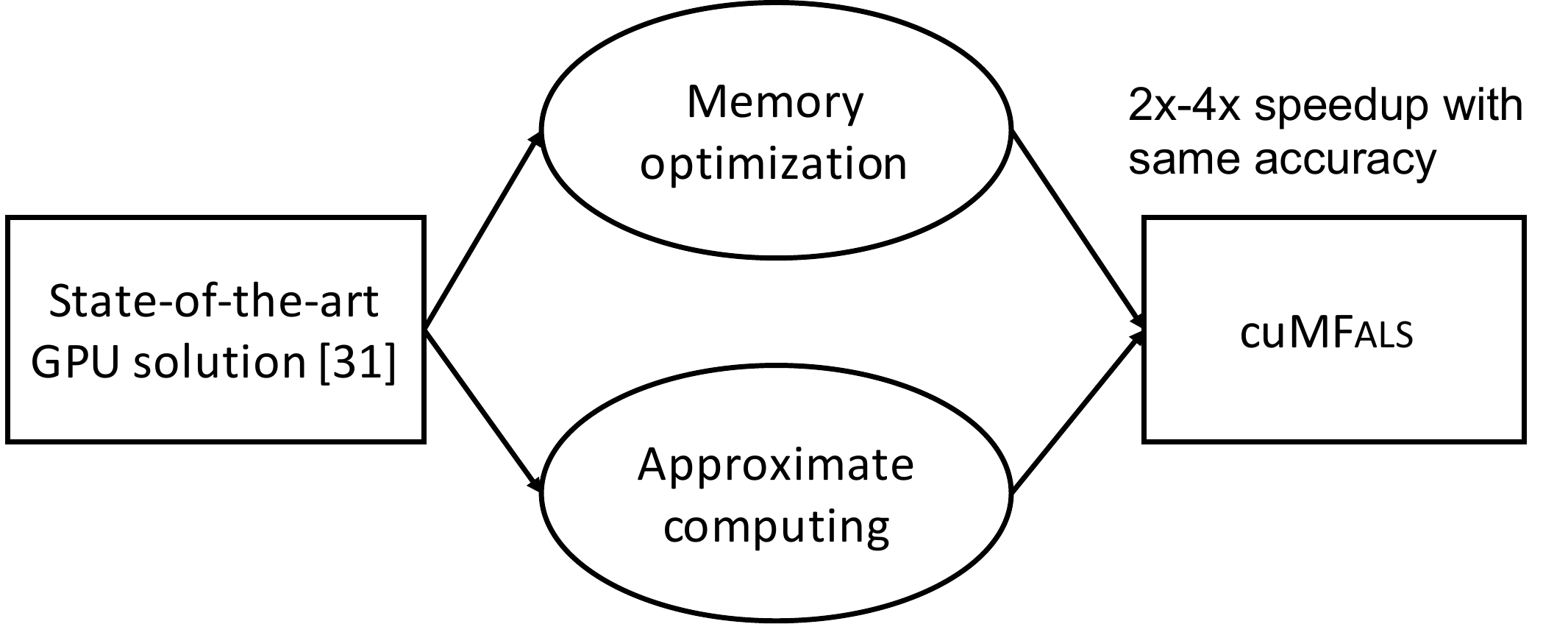}}
 \caption{We optimize the state-of-the-art GPU implementation via two directions: memory optimization and approximate computing. 
Combining the two, we achieve 2x - 4x speedup with the same accuracy.}
%\vspace{-0.5cm}
\label{fig:methodology}
\end{figure}

In this paper, we propose a novel approach in solving MF on GPUs, termed \algname, with major contributions in two-fold:
\begin{itemize}[leftmargin=*]
\item To fully utilize the GPU memory hierarchy, we identify hotspot variables to retain data as close to compute as possible. Afterward, the memory loading process is accelerated through an innovative and non-conventional scheme by exploiting GPU architectural features such as occupancy and cache.
\item We develop an iterative conjugate gradient (CG) solver on GPUs. This approximate solver reduced the compute complexity from $\mathcal O(f^3)$ to $\mathcal O(f^2)$ ($f$ is the dimension of latent features) without hurting convergence. This optimization brings a speedup of \textbf{4x} compared to calculating in exact. Moreover, CG works naturally with Nvidia's newly developed half precision feature, which further doubles the speed. 
\end{itemize}
A graphical illustration of our proposed approach is shown in Figure \ref{fig:methodology}. By jointly optimizing the memory loading scheme and the approximate compute strategy, we are able to not only outperform all distributed CPU solutions by a large margin, but also make \textbf{2x-4x} improvements over the state-of-the-art GPU implementation. Such performance gains have been validated through extensive experiments on various GPU architectures. Our implementation is open-sourced on GitHub\footnote{\url{https://github.com/cuMF/cumf_als}}, available as a library to accelerate many applications.  
%The former exploits GPU memory hierarchy to increase data reuse, while %the later reduces unnecessary computing without hurting convergence. 

\section{Preliminaries}
\begin{table*}[t]
 \caption{\textbf{Compute and memory complexity per epoch: ALS vs. SGD.} ALS is compute intensive and SGD is memory intensive, so they need different optimizations on GPUs.}
\label{tbl:complexity}
\centering
\begin{tabular}
{|p{0.05\textwidth}|p{0.1\textwidth}|p{0.15\textwidth}|p{0.15\textwidth}|p{0.035\textwidth}|}  \hline
\multicolumn{2}{|c|}{}& \textbf{Compute (C)} & \textbf{Memory (M)} & \textbf{C/M}\\ \hline
\multirow{2}{*}{\textbf{ALS}} & get\_hermitian & $\mathcal O(N_zf^2)$& ${\mathcal O(N_zf+(m+n)f^2)}$ & $f$\\ 
&solve & $\mathcal O((m+n)f^3)$& $\mathcal O((m+n)f^2)$& $f$ \\ \hline
\multicolumn{2}{|c|}{\textbf{SGD}}& $\mathcal O(N_zf)$& $\mathcal O(N_zf)$ & 1\\ \hline
\end{tabular}
%\vspace{-0.3cm}
\end{table*}
\label{sec:pre}
Matrix factorization (MF) factorizes a matrix $R \in \mathbb R^{m\times n}$ (with $N_z$ non-zero elements) into two low-rank matrices $X \in \mathbb R^{m\times f}$ and $\Theta \in \mathbb R^{n\times f}$, such that $R\approx X \cdot \Theta^{T}$. For any $u$ and $v$, such tat $1\leq u\leq m$ and $1\leq v\leq n$, $r_{uv}$ is the $(i,j)$ entry of $R$. Thus, $r_{uv}\approx \boldsymbol{x}_u^T \cdot \boldsymbol\theta_v$, where $\boldsymbol{x}_u, \boldsymbol{\theta}_v\in \mathbb R^f$ are the $u^{th}$ column of $X^T$ and the $v^{th}$ column of $\Theta^T$, respectively. Then the optimization problem of MF is given as:
\begin{equation}
\small
 \min_{X, \Theta} \sum\limits_{r_{uv}\neq0} (r_{uv} - \boldsymbol{x}_u^T\boldsymbol{\theta}_v)^2
 +\lambda (\sum_{u}n_{x_u}||\boldsymbol{x}_u||^2 +\sum_{v}n_{\theta_v}||\boldsymbol{\theta}_v||^2),
\label{eq-mf}
\end{equation}
where $n_{x_u}$ and $n_{\theta_v}$ are the number of non-zero elements of $\boldsymbol{x}_u$ and $\boldsymbol{\theta}_v$, respectively; $\lambda$ is the regularization parameter.  Two important approaches ALS and SGD both minimize equation (\ref{eq-mf}), yet using different approaches that we will discuss the next. \\

%%%%%%%%%%%%%%%%%%%%%%%%%%%%%%%%%%%%%%%%%%%%%%%%%%%%%%%%%%%%%%%%%%%%%%%%%%%%%%
%\vspace*{-0.1in}
\noindent {\bf ALS}: ALS is an iterative method that first optimizes $X$ while fixing $\Theta$, and then solves $\Theta$ while fixing $X$.  In every iteration, all observations ($r_{uv}\neq0$) are used to update the current variable.  Moreover, both subproblems are convex and the update procedures for them are given below. 

\vspace*{0.05in}
\noindent \textbf{Update $X$}: The optimal solution of the $u^{th}$ column of $X^T$ is obtained by solving the following linear system:
\begin{equation}
\small
\label{als-x}
 \sum\limits_{r_{uv}\neq0} (\boldsymbol\theta_v \boldsymbol{\theta}_v^T+\lambda I) \cdot \boldsymbol{x}_u  = \Theta^T\cdot R_{u*}^T.
\end{equation}

\noindent \textbf{\bf Update $\Theta$}: Similarly, the optimal solution of $v^{th}$ column of $\Theta^T$ is obtained by solving: 
\begin{equation}
\small
\label{als-theta}
\sum\limits_{r_{uv}\neq0} (\boldsymbol{x}_u \boldsymbol{x}_u^T+\lambda I) \cdot \boldsymbol\theta_v = X^T\cdot R_{*v}.
\end{equation}
Here, $R_{u*}$ and $R_{*v}$ are the $u^{th}$ row and $v^{th}$ column of $R$, respectively.  It is worth mentioning that the updates of each $\boldsymbol{x}_u$ and $\boldsymbol\theta_v$ are independent.  In other words, every row of matrix $X$ can be updated in parallel while keeping $\Theta$ fixed. The same procedure is applicable to update $\Theta$ as well.  To ease repeating, throughout the rest of this paper, we will only focus on solving $X$. 

The solution of equation (\ref{als-x}) has a closed form as 
\begin{equation}
\small
\boldsymbol{x}_u  = (\sum\limits_{r_{uv}\neq0} \boldsymbol\theta_v \boldsymbol{\theta}_v^T+\lambda I)^{-1} \cdot \Theta^T\cdot R_{u*}^T,
\end{equation}
which involves calculating a matrix inverse. Nevertheless, matrix inverse is compute intensive and unnecessary in solving the linear system in (\ref{als-x}).  Instead, many literatures \cite{sparkmllib16,als-10} solve the problem in a two-step fashion: 

(i) compute intermediate results of $A_u=\sum\limits_{r_{uv}\neq0}(\boldsymbol\theta_v \boldsymbol\theta_v^T+\lambda I)$ and $b_u = \Theta^T\cdot R_{u*}^T$, which are called \texttt{get\_hermitian} and \texttt{get\_bias}, respectively; 

(ii) solve the linear system, which will be referred as \texttt{solve}.

Our method also follows the two-step solving scheme.  However, for each step, we propose a novel technique to better utilize both compute and memory resources of GPUs. Comparing \texttt{get\_hermitian} and \texttt{get\_bias}, we note that the compute complexity is dominated by the former one.  Thus, we firstly focus on the optimization of \texttt{get\_hermitian}, not \texttt{get\_bias} in ALS in this paper. \\

%%%%%%%%%%%%%%%%%%%%%%%%%%%%%%%%%%%%%%%%%%%%%%%%%%%%%%%%%%%%%%%%%%%%%%%%%%%%%%
%\vspace*{-0.08in}
\noindent {\bf SGD}: SGD is also an iterative algorithm.  However, differ to ALS, under each iteration, SGD only work with a small subset of observations denoted as $\Omega^{k}$ (\emph{a.k.a.} mini-batch), where $k$ is the number of iterations.  Usually, samples in $\Omega^{k}$ are randomly selected from all observations.  Then, the updating equations for both $X$ and $\Theta$ for the $k^{th}$ iteration are given as:
\begin{equation}
\small
\label{eq-sgd}
\begin{aligned}
\boldsymbol{x}_u^{k} & = \boldsymbol{x}_u - \alpha^{k} \sum_{v: r_{u,v} \in \Omega^{k}}(\boldsymbol{x}_u^T \boldsymbol\theta_v - r_{uv}) \boldsymbol\theta_v + \lambda \boldsymbol{x}_u, ~\text{and}
\\
\boldsymbol\theta_v^{k} & = \boldsymbol\theta_v - \alpha^{k} \sum_{u: r_{u,v} \in \Omega^{k}} (\boldsymbol{x}_u^T \boldsymbol\theta_v - r_{uv})\boldsymbol{x}_u + \lambda \boldsymbol\theta_v,
\end{aligned}
\end{equation}
where $\alpha^{k}$ is known as the learning rate.  The vanilla SGD algorithm requires passing over randomly sampled data multiple times (till converge).  When multiple updates run in parallel, for example, two samples $r_{uv}$ and $r_{uv'}$ are updating at the same time, their updates to $\boldsymbol{x}_u$ may overwrite each other. To address this issue, previous studies either partition $R$ into blocks with no overlapping rows and columns~\cite{libmf-13, nomad14, DSGD-kdd11, DBLP:conf/icdm/TeflioudiMG12}, or let multiple workers independently update ignoring conflicts \cite{hogwild-nips11}. \\

%%%%%%%%%%%%%%%%%%%%%%%%%%%%%%%%%%%%%%%%%%%%%%%%%%%%%%%%%%%%%%%%%%%%%%%%%%%%%%

\noindent {\bf Complexity}: Following the roofline model~\cite{roofline2009}, we calculate the computation and memory complexity for both ALS and SGD and summarize them in Table~\ref{tbl:complexity}. Comparatively, ALS has a higher compute-to-memory ratio than SGD, which means ALS is compute intensive while SGD is memory intensive. Although the compute complexity of ALS is heavier than SGD per iteration, the number of iterations to converge is significant fewer \cite{mf-computer09}.  Moreover, parallelization of ALS is easier since no sophisticated locking scheme is needed \cite{mf-computer09,cumfsgd}.  At last, ALS is more suitable for the case of MF with implicit inputs, which makes it more broadly applicable. 

Based on this complexity analysis, we focus our study in accelerating ALS in this paper, while SGD would be used as comparing topic.

\vspace*{0.1in}
\noindent {\bf Approximate computing}: This term applied to computation that returns approximated result as the trade-off between accurracy and cost/performance. \cite{Esmaeilzadeh12} explores some of the applications and hardware designs for approximate computing. Their work showed acceleration gain of 1.9X to 2.1X with only 2.5\% quality loss. \cite{Mittal16} provides a detailed survey on both software techniques and hardware design for a large variety of applications to leverage the power of approximate computing for cost or performance gain. We exploit approximate computing in two aspects.  Firstly, our iterative process in linear equation solver would stop within some tolerable converaging values.  Secondly, we use reduced precison hardware feature to maximize the utilization of memory bandwidth and memory capacity.

\section{Memory optimization for high flops}
\label{sec:memory}
As mentioned in section~\ref{sec:pre}, an ALS update includes two steps, \textit{i.e.} \texttt{get\_hermitian} and \texttt{solve}. This section describes the memory optimization on \texttt{get\_hermitian} and the next section introduces the approximate computing techniques on \texttt{solve}. As seen from Table~\ref{tbl:complexity}, \texttt{get\_hermitian} has a compute complexity of $\mathcal O(N_zf^2)$. This is big in large-scale problems where $N_z$ can be tens of billions, which leads to our first observation.

\vspace*{0.05in}
\noindent\textbf{Observation 1. \texttt{get\_hermitian} is compute intensive. }
\vspace*{0.05in}

\noindent For a compute intensive function to achieve high FLOPS, it needs to retain data as close as possible to compute units~\cite{hennessy2011computer,roofline2009}. In other words, it needs effective caching to reduce read from external memory, \textit{a.k.a.} DRAM, as DRAM cannot sustain high FLOPS of GPUs. Specifically to \algname, we need to identify the frequently-used variables, exploit the GPU memory hierarchy and place hotter variables in faster memory. This leads to the following solution.

\vspace*{0.05in}
\noindent\textbf{Solution 1. Utiilize register and shared memory.}
\vspace*{0.05in}

\noindent To decide what to cache and to where, we analyze the memory usage in calculating $A_u$:
\begin{itemize}[leftmargin=*]
\item $A_u$ is read and written once when adding each $\boldsymbol\theta_v\boldsymbol\theta_v^T$. Therefore, $A_u$ is read and written by $n_{x_u}$ times, that is $N_z/m$ on the average.
\item Each $\boldsymbol\theta_v$ needs to be read $f$ times when calculating $\boldsymbol\theta_v\boldsymbol\theta_v^T$. 
\end{itemize}

\begin{figure*}[t]
\center{\includegraphics[width=0.8\linewidth]
        {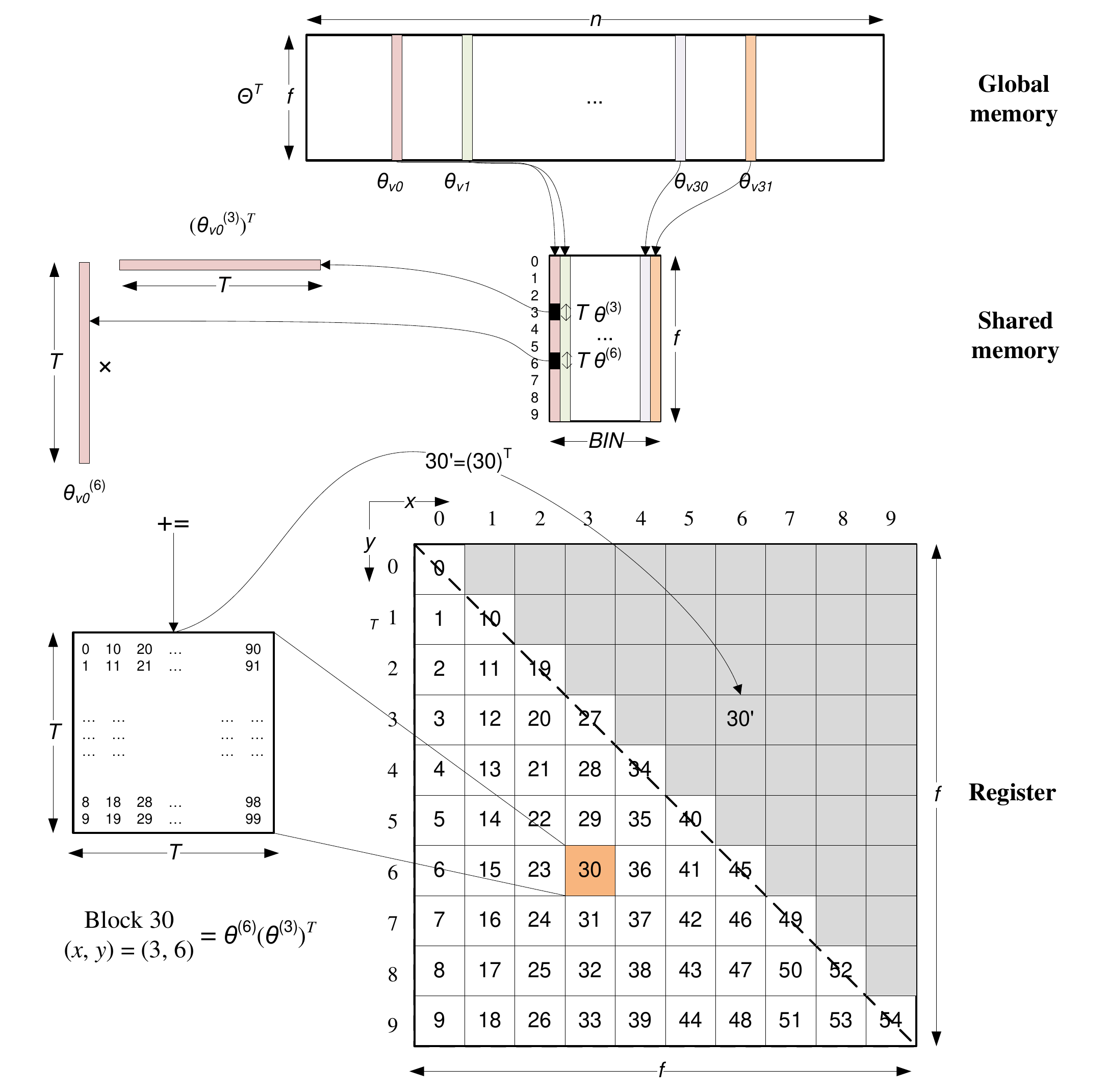}}
 \caption{\textbf{The memory optimization to} \texttt{get\_hermitian}. For a given $\boldsymbol{x}_u$, its required $\boldsymbol\theta_v$s such that $r_{uv}\neq 0$, are staged from $\Theta^T$ from global memory to a shared memory buffer of size $BIN*f$, in batches. For each $\boldsymbol\theta_v$ in shared memory, we calculate $\boldsymbol\theta_v\boldsymbol\theta_v^T$ in tiles of size $T$, and add to sub-blocks of $A_u$ in registers. Each sub-block in $A_u$ adds the outer product of two tiles in $\boldsymbol\theta_v$. Consider the symmetricity, we only calculate tiles with coordinates of $x\leq y$. For example, for $\boldsymbol\theta_{v0}$, tiles 3 and 6 need to do outer product and add to sub-block 30 in $A_u$, i.e., $block_{30}\mathrel{+}= \boldsymbol\theta_{v0}^{(6)} {(\boldsymbol\theta_{v0}^{(3)})}^T$, and $block_{30'}={(block_{30})}^T$. $A_u$ in registers is flushed to global memory after all required $\boldsymbol\theta_v\boldsymbol\theta_v^T$s are added into it.}
\label{fig:kernel}
\end{figure*}

We can now allocate variables into different places in GPU memory hierarchy based on their reuse. Usually $N_z/m\gg f$, and as a result $A_u$ is more frequently accessed than $\boldsymbol\theta_v$.  Therefore, $A_u$ deserves the fastest cache, \textit{i.e.} register, and $\boldsymbol\theta_v$ is put into shared memory, the second fastest cache. Figure~\ref{fig:kernel} illustrates the memory optimization to \texttt{get\_hermitian}. For a given $\boldsymbol{x}_u$, its required features, \textit{i.e.} $\boldsymbol\theta_v$s such that $r_{uv}\neq 0$, are staged from $\Theta^T$ in global memory (the matrix at the top) into a shared memory space of size $BIN\times f$ (the thinner matrix in the middle), in batches. For each staged feature $\boldsymbol\theta_v$, we calculate $\boldsymbol\theta_v\boldsymbol\theta_v^T$ in tiles of size $T$ and add to the corresponding sub-blocks of $A_u$ in registers (the symmetric matrix at the bottom). Each sub-block in $A_u$ aggregates the outer product of two tiles in $\boldsymbol\theta_v$. Consider the symmetricity of $A_u$, we only need to calculate the bottom half of it. $A_u$ stored in registers is flushed to global memory when all required $\boldsymbol\theta_v\boldsymbol\theta_v^T$s are added in.

Because we choose to excessively use registers, they become the constrained resources. Consequently, the occupancy of \texttt{get\_hermitian}, \textit{i.e} number of $A_u$s that can be calculated concurrently is low. 

\vspace*{0.05in}
\noindent\textbf{Observation 2. Aggressive use of registers leads to low occupancy, which makes read from global memory latency-bound instead of bandwidth-bound.}
\vspace*{0.05in}

\noindent Current Nvidia GPUs have 65536 float registers in each stream-multiprocessor (SM). When $f=100$, each thread of \texttt{get\_hermitian} needs 168 registers and each block needs 64 threads. As a result, an SM can hold $65536/(168 \times 64)\approx 6$ thread-blocks, i.e. an SM can update 6 rows concurrently. Compared with the SM capacity to hold 32 thread-blocks, this is a low occupancy. This indicates that there are relatively few concurrent threads loading from global memory, which leads to the next solution.

\vspace*{0.05in}
\noindent\textbf{Solution 2. Use non-coalesced and cache-assisted read, which is non-conventional but proven faster.}  
\vspace*{0.05in}

\noindent In GPU programming, memory coalescing is considered as a best practice to achieve good performance~\cite{kirk2012programming}. Memory coalescing means that adjacent threads should access adjacent global memory addresses. It can consolidate memory load requests and avoid wasting the bandwidth. When using coalescing, read from global memory can bypass L1 cache, because loaded data are all used.
%High occupancy means that there should be sufficient threads on-the-fly, so that compute and memory units are fully utilized. 

\begin{figure}[t]
\center{\includegraphics[width=\linewidth]
        {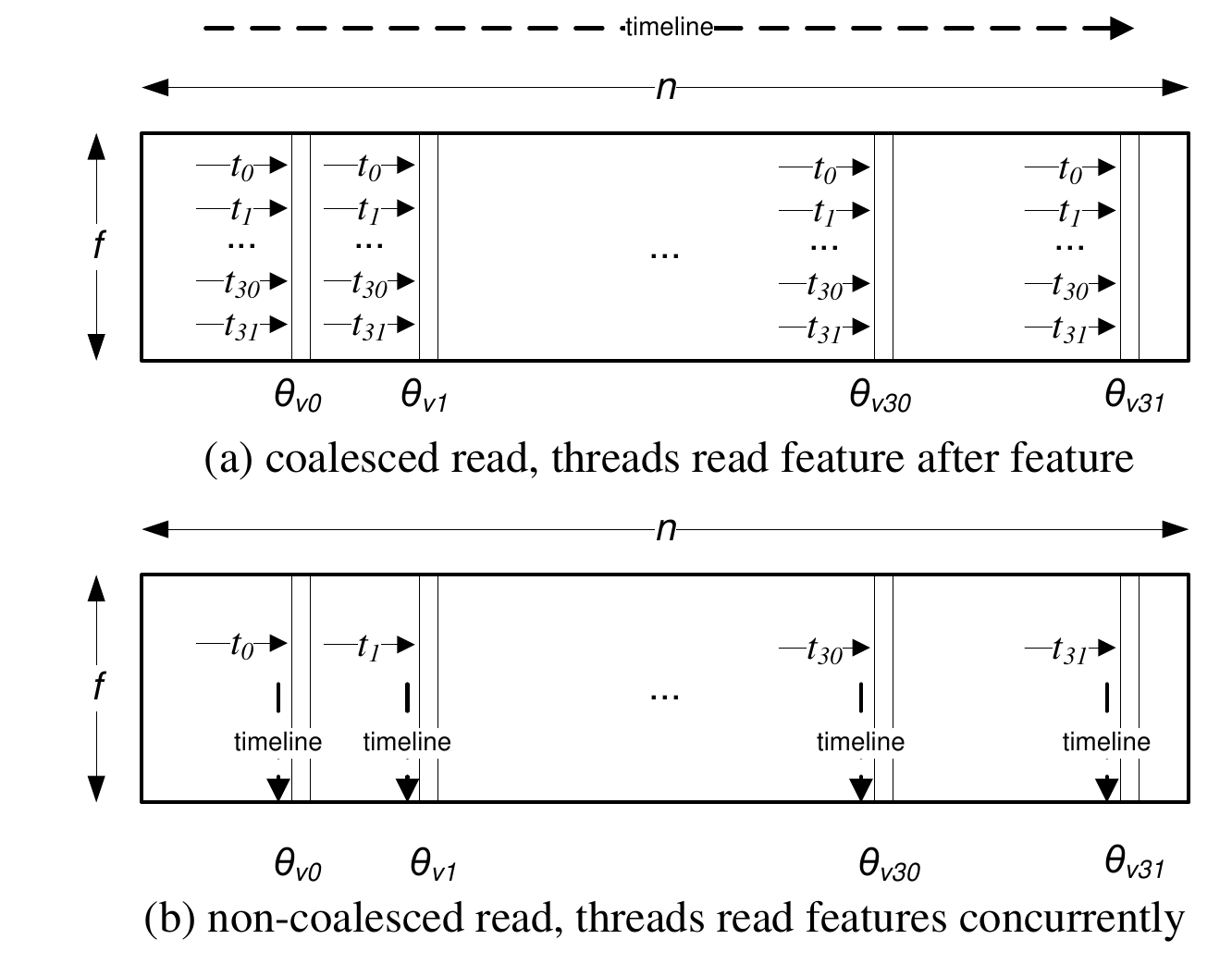}}
 \caption{\textbf{Load from global to shared memory in} \texttt{get\_hermitian}. (a) coalesced read: all threads read one column before moving to the next. This issues fewer memory instructions but is lack of parallelism. (b) Non-coalesced read: multiple threads read multiple columns concurrently. In low occupancy, the columns are cached and subsequent non-coalesced reads will hit cache.}
\label{fig:coalesce}
\end{figure}

Given the low occupancy of \texttt{get\_hermitian}, coalesced read, despite its efficiency, cannot saturate the memory bandwidth. On the other hand, when the occupancy is low, the working data can almost fit into L1 cache. For example, when $f=100$ and $BIN=32$, the $\boldsymbol\theta_v$s being actively loaded per SM is $100\times 32\times 6$ (thread-block)$\times 4$ (bytes per float)$=75$ KB. This number is between Nvidia Maxwell's L1 cache of 48 KB and L2 cache of 128 KB (3 MB shared by 24 SMs). Inspired by this observation, we use a parallel but non-coalesced read scheme as illustrated in Figure~\ref{fig:coalesce} (b). Without losing generality, we load 32 features $\boldsymbol\theta_{v0}, \boldsymbol\theta_{v1}, \boldsymbol\theta_{v30},\dots ,\boldsymbol\theta_{v31}$ using 32 threads. With the coalesced scheme in Figure~\ref{fig:coalesce} (a), 32 threads together read one $\boldsymbol\theta_{v}$ column before moving to the next one. Alternatively, in the non-coalesced scheme in Figure~\ref{fig:coalesce} (b), 32 threads read 32 columns concurrently, with each thread reading one column. Because of the small working data set size, L1 and L2 cache can efficiently serve as the coalescing cache. That is, the non-coalesced load requests issued from $t_0, t_1, \dots, t_{30}, t_{31}$ are going to hit L1 and L2, which makes it even more efficient than coalesced read.

\begin{figure}[t]
\center{\includegraphics[width=\linewidth]
 {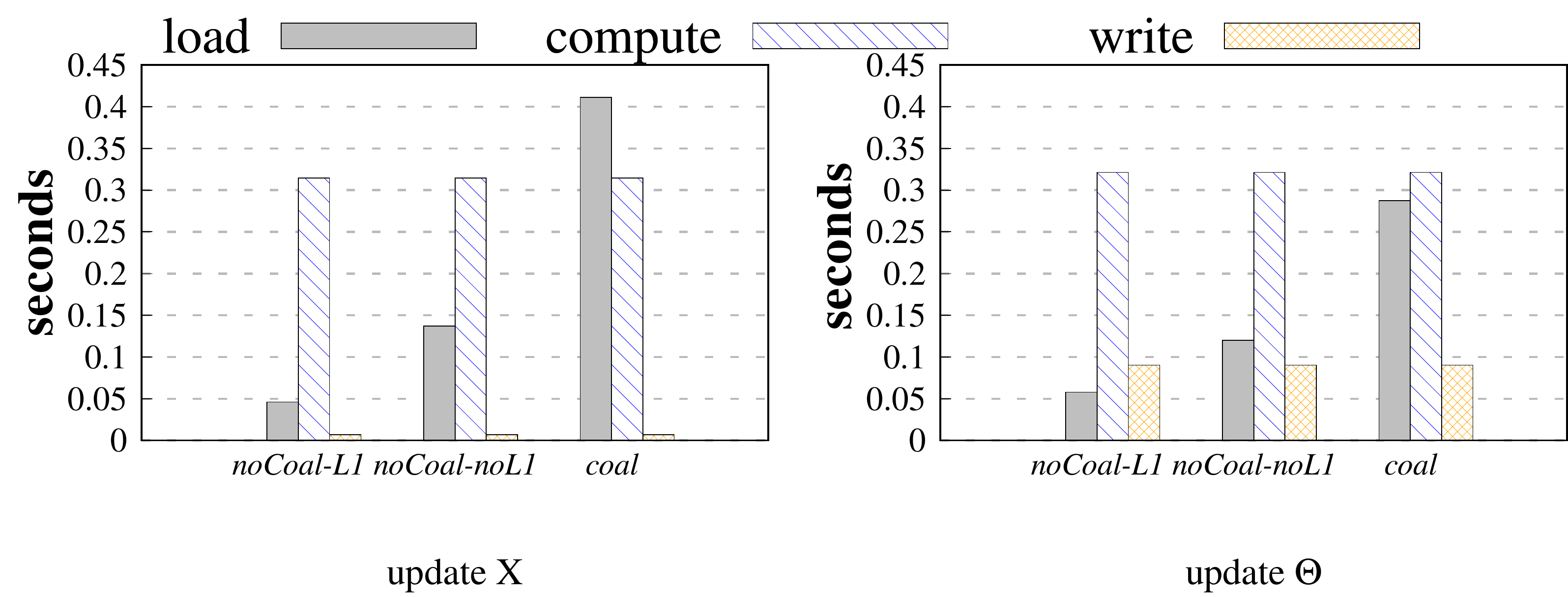}}
 \caption{\textbf{The performance of coalesced and non-coalesced read from global to shared memory in} \texttt{get\_hermitian}, using the Netflix dataset. Bar \texttt{load} shows the memory load time; non-coalesced read with L1 cache (\texttt{nonCoal-L1}) is the fastest.
 %Non-coalesced read is faster because in such low occupancy, memory load is latency- (not bandwidth-) bounded and working data set fits in cache.
 }
\label{fig:hermitian-perf}
\end{figure}

To showcase the effectiveness of solution 2, we measure the performance of coalesced and non-coalesced read in \texttt{get\_hermitian}. We use the Netflix dataset (see Section~\ref{sec:datasets} for more details on datasets) and measure the time of three phases in \texttt{get\_hermitian}: load from global memory to shared memory (\textbf{load}), compute $A_u$ (\textbf{compute}), and write $A_u$ to global memory (\textbf{write}). Figure~\ref{fig:hermitian-perf} shows the performance of both update-X and update-$\Theta$ procedures, in three different settings: \texttt{coal} means coalesced read, the setting illustrated in Figure~\ref{fig:coalesce} (a); \texttt{nonCoal-L1} means non-coalesced read, the setting in Figure~\ref{fig:coalesce} (b) and with L1 cache; \texttt{nonCoal-noL1} means non-coalesced read with L1 cache bypassed. Result shows: 
\begin{itemize}[leftmargin=*]
\item For shared memory load, non-coalesced with L1 performs best; non-coalesced without L1 is worse and coalesced read is worst.
\item The compute time is almost constant in all settings. This is because they all need $N_z\times f^2$ fused multiple-add (FMA) operations.
\item Update-X and update-$\Theta$ need to write $m\times f^2$ and $n\times f^2$ floats to global memory, respectively. Since $m<n$ in Netflix data, update-$\Theta$ takes longer in \texttt{write}.
\end{itemize}

%\textcolor{red}{TODO: move Sec~\ref{sec:exp-flops} here?}

%Also, we are not constrained by coalesced write.
\section{Approximate computing in solver}
Section~\ref{sec:memory} describes how to efficiently obtain $A_u$. After that, we need to solve $m$ equations $A_u\boldsymbol{x}_u=\boldsymbol{b}_u$ as illustrated in equation~\eqref{als-x}. 
\label{sec:solver}
\subsection{Approximate solver with CG}
The direct solver, \textit{e.g.}, the batch LU solver in cuBLAS~\cite{cublas}, gives an exact solution to $A\boldsymbol{x}=\boldsymbol{b}$ with compute complexity of $\mathcal O(f^3)$, or $\mathcal O(m\times f^3)$ for \textit{m} rows. This cubic complexity leads to long solve time, especially when $m$ is big. As shown in Table~\ref{tbl:complexity}, when $m$ becomes big, $R$'s rows become sparse, and $m\times f^3$ gets closer to $N_z\times f^2$. To demonstrate this, we measure the solver time of 10 ALS iterations on Netflix data. Column \texttt{LU\_FP32} in Figure~\ref{fig:solver} shows that, the time taken by the LU solver is almost twice as much as that by \texttt{get\_hermitian}. This clearly indicates that after applying optimization on \texttt{get\_hermitian}, \texttt{solve} executing time now becomes dominant. This leads to the following observation.

\vspace*{0.05in}
\noindent\textbf{Observation 3. Solve is compute intensive and dominant.}
\vspace*{0.05in}

\noindent This observation inspires us to seek an alternative to the direct solver. We notice that, as an iterative approach, ALS updates $X$ and $\Theta$ based on estimations from the previous iteration. As errors exist in estimations, the solution of each step is inherently inaccurate. Therefore, solution accuracy may be sacrificed in exchange for compute speed, leading to our attempt for an approximate solver.

\vspace*{0.05in}
\noindent\textbf{Solution 3. An approximate conjugate gradient solver.}
\vspace*{0.05in}

\noindent The iterative CG solver is introduced in ~\cite{hestenes1952methods}. With $f$ iterations each of complexity $\mathcal{O}(f^2)$, it yields the exact solution with complexity $\mathcal{O}(f^3)$. Based on this, we seek to further reduce computation while maintaining convergence quality. The pseudo code of our approximate CG solver is summarized in Algorithm~\ref{alg:cg}, where $f_s$ is given to control the number of iterations (see Line 3), and $\epsilon$ for tolerance control. Empirically this approximation does not impact ALS's convergence, while effectively reducing the solver's complexity from $\mathcal{O}(f^3)$ to $\mathcal{O}(f^2)$ when $f_s\ll f$. 

\begin{algorithm}[t]
\caption{The CG solver for $A\boldsymbol{x}=\boldsymbol{b}$.}
\label{alg:cg}
\begin{algorithmic}[1]
\Procedure{CGSolve}{$A,\boldsymbol{x},\boldsymbol{b},f_s,\epsilon$}
\State $\boldsymbol{r}=\boldsymbol{b}-A \cdot \boldsymbol{x}$; \quad $\boldsymbol{p}=\boldsymbol{r}$; \quad  $rs_{old}=\boldsymbol{r}^T\cdot \boldsymbol{r}$ 
\For{$j=1:f_s$}
\State $\boldsymbol{a}_p=A\cdot \boldsymbol{p}$;
\quad $\alpha=rs_{old}/(\boldsymbol{p}^T \cdot \boldsymbol{a}_p)$
\State $\boldsymbol{x}=\boldsymbol{x}+\alpha\boldsymbol{p}$;
\quad $\boldsymbol{r}=\boldsymbol{r}-\alpha\boldsymbol{p}$
\State $rs_{new}=\boldsymbol{r}^T\cdot \boldsymbol{r}$
\If {$\sqrt{rs_{new}}<\epsilon$}
    \State \textbf{break}
\EndIf
\State $\boldsymbol{p}=\boldsymbol{r}+(rs_{new}/rs_{old})\boldsymbol{p}$
\State $rs_{old}=rs_{new}$
\EndFor
\State \Return $\boldsymbol{x}$
\EndProcedure
\end{algorithmic}
\end{algorithm}

% \begin{figure}
% \center{\includegraphics[width=0.8\linewidth]
%         {figs/approximate.pdf}}
%  \caption{\textbf{Exact and approximate solver in ALS.}}
% \label{fig:approximate}
% \end{figure}
%%%%%%%%%%%%%%%%%%%%%%%%%%%%%%%%%%%%%%%%%%%%%%%%%%%%%%%%%%%%
%%%%%%%%%%%%%%%%%%%%%%%%%%%%%%%%%%%%%%%%%%%%%%%%%%%%%%%%%%%%
\subsection{Use reduced precision}
Replacing LU solver with an approximate CG solver, the solver's compute-to-memory ratio (recall Table~\ref{tbl:complexity}) now drops from $\mathcal{O}(f)$ to $\mathcal{O}(1)$, converting the original compute intensive problem into a memory intensive one.

\vspace*{0.05in}
\noindent\textbf{Observation 4. CG solver is memory intensive.}
\vspace*{0.05in}

\noindent 
%Reading $f^2$ elements of $A$ makes the memory complexity of solving $A\boldsymbol{x}=\boldsymbol{b}$ to be $\mathcal{O}(f^2)$. The compute complexity of CG solver is $\mathcal{O}(f^2)$ compared with LU solver's $\mathcal{O}(f^3)$. Consequently, the CG solver introduced in solution 3 makes step \texttt{solve} memory intensive.
As seen in Algorithm~\ref{alg:cg}, CG solver is dominated by dense matrix-vector multiply $A\cdot \boldsymbol{p}$ (Line 4), which is in turn dominated by reading $A$ that is of memory complexity $\mathcal{O}(f^2)$. This insight inspires us that further acceleration is possible by reducing the size of $A$.

\vspace*{0.05in}
\noindent\textbf{Solution 4. Use reduced precision in CG solver to double the effective memory bandwidth.}
\vspace*{0.05in}

\begin{figure}[t]
\center{\includegraphics[width=0.75\linewidth]
        {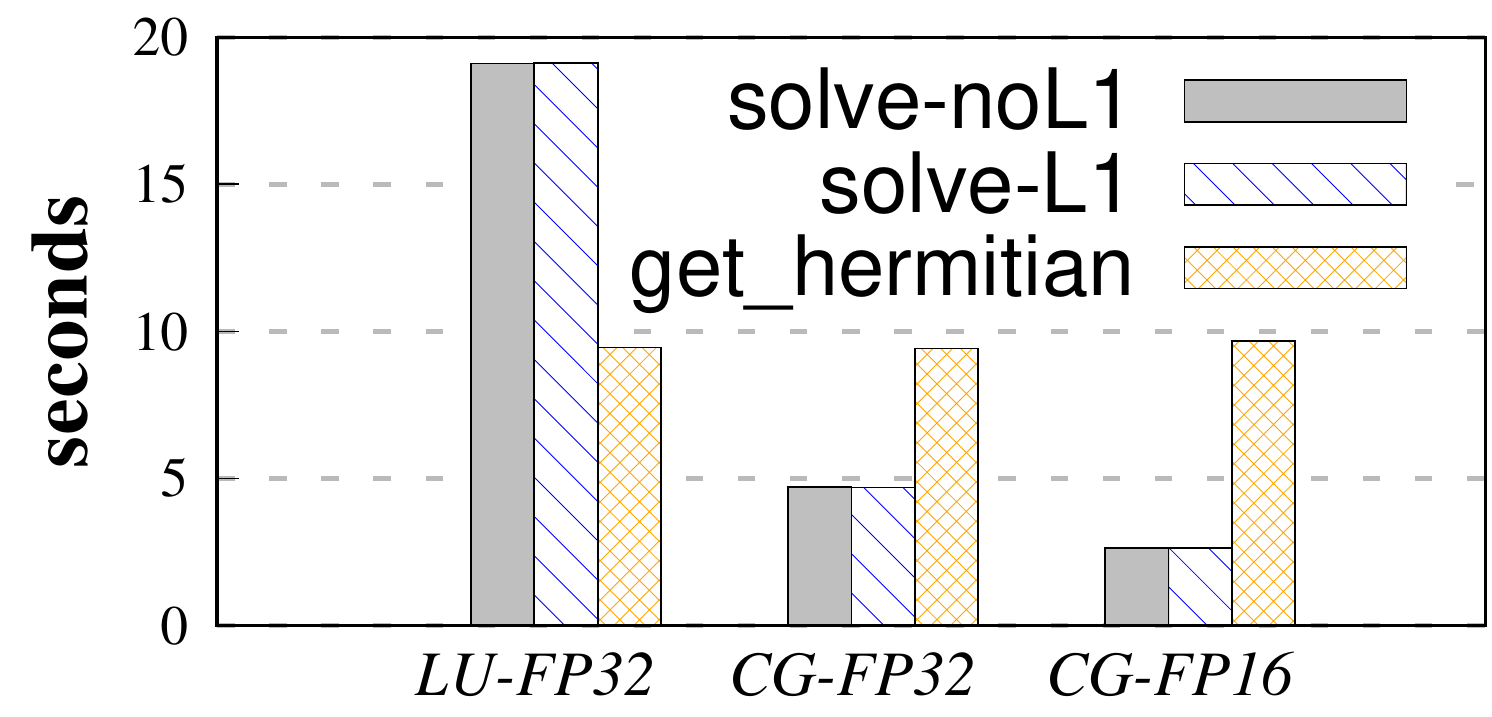}}
 \caption{The solver time of 10 ALS iterations using Netflix data on Nvidia Maxwell Titan X. $f=100$ and we use $f_s=6$ (the smallest number that does not hurt convergence) for CG. CG-FP32 is $1/4$ of the LU-FP32 time; CG-FP16 takes $1/2$ of the CG-FP32 time. Using L1 (\texttt{solve-L1}) takes the same time as without it (\texttt{solve-noL1}).}
\label{fig:solver}
\end{figure}

\noindent We choose the newly introduced 16-bit floating point format (FP16, compared with the default 32-bit floating point format FP32) in Nvidia GPUs to store $A$. This optimization saves 50\% memory bandwidth to load $A$ and consequently doubles the loading speed. To validate, we run 10 iterations of ALS using Netflix data on a Maxwell GPU. Figure~\ref{fig:solver} shows the total solver time. The time of CG solver with FP32 (\texttt{CG-FP32}) is only $1/4$ of that of LU solver with FP 32 (\texttt{LU-FP32}). When CG uses FP16, \texttt{CG-FP16} takes $1/2$ of the time compared with \texttt{CG-FP32}. In total, CG-FP16 can reduce the run-time to $1/8$ compared with LU-FP32.
 
%A caveat here is that, FP16 in CUDA has a range of $[5.96\times10^{-8},6.55\times10^4]$. As a result, we need to normalize $R$ such that $A$ fits in the range of FP16.

\vspace*{0.05in}
\noindent\textbf{Does L1 cache benefit the CG solver?}
\vspace*{0.05in}

\noindent Figure~\ref{fig:solver} also illustrates that, loading $A_u$ with L1 cache does not yield any performance benefit. This is coherent with the analysis in section~\ref{sec:memory}: L1 cache is only useful to coalesce the non-coalesced memory access when occupancy is low. With batch CG's high occupancy and coalesced read, L1 cache is not useful at all. This also explains why L1 cache is disabled by default in Nvidia GPUs.
\section{Experiments}
%%%%%%%%%%%%%%%%%%%%%%%%%%%%%%%%%%%%%%%%%%%%%%%%%%%%%%%%%%%%%%%%%%%%%%%%%%%%%%
In this section, we show the advantages of the proposed \algname framework compared to a set of state-of-the-art implementations for both CPU and GPU. Our experiments are designed to answer the following questions: 
\begin{itemize}[leftmargin=*]
  \item How fast \algname is compared to competing implementations?
  \item How efficiently \algname utilizes compute resource (in terms of FLOPS) and memory bandwidth of GPU, as argued in sections~\ref{sec:memory} and~\ref{sec:solver}? 
    \item How does \algname compare to SGD? 
    \item Can \algname extent to the setting of MF with implicit feedback (\emph{a.k.a} one-class or positive-unlabeled inputs)?
\end{itemize}
%%%%%%%%%%%%%%%%%%%%%%%%%%%%%%%%%%%%%%%%%%%%%%%%%%%%
\subsection{Datasets}
\label{sec:datasets}
We utilize three publicly available datasets as follows: 
\begin{itemize}[leftmargin=*]
\item \textit{\textbf{Netflix}} \cite{netflix08}: The Netflix dataset consists ratings on movies.  Each rating is in the scale of one to five.  
\item \textit{\textbf{YahooMusic}} \cite{kdd-cup-yahoomusic-11}: Similar to the Netflix, this dataset contains 250 million ratings in the range of 1 to 100 for music collected by the Yahoo! Music Radio service. 
\item \textit{\textbf{Hugewiki}} \cite{libmf-13}: Hugewiki contains a snapshot of Wikipedia. The observation matrix $R$ describes the frequency of English terms appeared in different documents.
\end{itemize}

The detailed statistics are summarized in Table~\ref{tbl:dataset}. We choose 0.92, 22 and 0.52 as the convergence value for Netflix, YahooMusic and Hugewiki, respectively, because these values are used by many papers and considered well-accepted.

\begin{table}[t]
\centering
\caption{\textbf{Benchmark datasets and parameters.}}
\label{tbl:dataset}
\begin{tabular}{|>{\centering\arraybackslash}m{1.4cm}|>{\centering\arraybackslash}m{1.2cm}|>{\centering\arraybackslash}m{0.9cm}|>{\centering\arraybackslash}m{0.8cm}|>{\centering\arraybackslash}m{0.5cm}|>{\centering\arraybackslash}m{0.5cm}|>{\centering\arraybackslash}m{0.65cm}|} \hline
\textbf{Dataset}	& \textbf{$m$} & \textbf{$n$}	& \textbf{$N_z$} & \textbf{$f$} &\textbf{$\lambda$} &\textbf{$RSME$} \\\hline
Netflix &480,189&17,770&99M&100&0.05&0.92\\ \hline
YahooMusic&1,000,990&624,961&252.8M&100&1.4&22\\ \hline
Hugewiki&50,082,603&39,780&3.1B&100&0.05&0.52\\ \hline
\end{tabular}
\end{table}

%%%%%%%%%%%%%%%%%%%%%%%%%%%%%%%%%%%%%%%%%%%%%%%%%%%%
%%%%%%%%%%%%%%%%%%%%%%%%%%%%%%%%%%%%%%%%%%%%%%%%%%%%
\subsection{Experiment setting}\label{sec:exp-config}
For the purpose of comprehensive evaluations, experiments are conducted on three different generations of Nvidia GPUs: Kepler, Maxwell and Pascal. Table~\ref{table:platform} illustrates the configurations of the three servers we use. CPU-only experiments are conducted on the most powerful Pascal server unless otherwise mentioned.  

For quantitative comparison, we follow the standard experiment setting \cite{libmf-13,nomad14} by reporting how fast the root mean square error (RMSE) on the testing test reduces.  The stopping criteria for all algorithms is when the RSME on testing set reaches an ``acceptable level''.   Specifically, the acceptable RSME is 0.92, 22.0 and 0.52 for Netflix, YahooMusic and Hugewiki, respectively.  Furthermore, we make use of the original training and testing files from the providers of Netflix and YahooMusic datasets while randomly extract 10\% of the data as the testing set for Hugewiki.  It is worth mentioning that, we focus on system-level efficiency in terms of running time instead of the recommendation accuracy. 
%To seek fair comparisons, we make sure all experiment environments are the same for all algorithms, except for their implementations.  
To achieve the goal, we use the same set of parameters ($f$ and $\lambda$) as reported from earlier works~\cite{hpdc2016,libmf-13,nomad14}, which is also shown in Table \ref{tbl:dataset}.

% \textbf{Hardware}. Experiments are conducted on three generations of Nvidia GPUs: Kepler, Maxwell and Pascal. Table~\ref{table:platform} shows the configuration of the three servers we use. CPU only experiments are done in the Pascal server unless otherwise mentioned. 

\begin{table}[t]
    \begin{center}
      \caption{\textbf{Config of Kepler, Maxwell and Pascal servers.}}
    \begin{tabular}{| >{\centering\arraybackslash}m{0.6 cm} | >{\centering\arraybackslash}m{7.2 cm} |}
    \hline
         \multicolumn{2}{|c|}{ \textbf{Kepler}} \\ \hline
     CPU & Two 8-core Intel Xeon E5-2667, 256 GB RAM\\ \hline
     GPU & Two Kepler K40, each: 4 TFLOPS, 12 GB RAM, 288 GB/s \\ \hline
     \multicolumn{2}{|c|}{ \textbf{Maxwell}} \\ \hline
     CPU & Two 12-core Intel Xeon E5-2670, 512 GB RAM \\ \hline
     GPU & Four Titan X, each: 7 TFLOPS, 12 GB RAM, 340 GB/s \\ \hline
  \multicolumn{2}{|c|}{\textbf{Pascal}} \\ \hline
   CPU & Two 10-core IBM Power8 with SMT 8, 512 GB RAM\\ \hline
     GPU & Four Tesla P100, each: 11 TFLOPS, 16 GB, 740 GB/s \\\hline
  \end{tabular}
  \label{table:platform}
  \end{center}
\end{table}

\subsection{Convergence speed: is \algname fast?}
\label{sec:exp-speed}

\begin{figure*}[t]
\centering
\includegraphics[scale=0.37,angle=0]{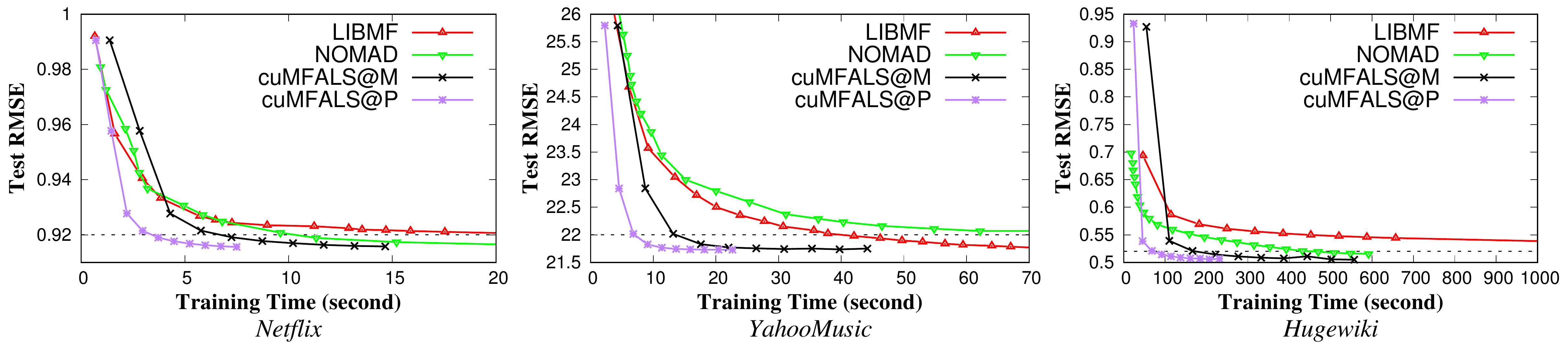}
\caption{\textbf{\algname vs. CPU solutions w.r.t. convergence time.} LIBMF uses 40 cores on one machine; NOMAD uses 32 machines for Netflix and YahooMusic, and 64 machines for Hugewiki. \algname uses one GPU for Netflix and YahooMusic, and four GPUs for Hugewiki. \algname on Maxwell (@M) and Pascal (@P) converges significantly faster than all other approaches.}\label{figure:compare}
\label{fig:speed}
\end{figure*}

There are many studies and systems on accelerating MF~\cite{factorbird14,libmf-13,libmf++,sparkler13,nomad14,DSGD-kdd11,ccd++-icdm12,hogwild-nips11,kdd15mf,cumfsgd,ccd++17}. Among them, we compare with the representative works below because \textbf{they are with state-of-art performance (i.e., convergence speed) or scalability}. 

\begin{itemize}[leftmargin=*]
\item
{\bf LIBMF}~\cite{libmf-13,libmf++}: The state-of-the-art CPU-based multi-thread solution using a single machine. 
\item
{\bf NOMAD}~\cite{nomad14}: NOMAD is a CPU-based solution using SGD. Different from LIBMF, it runs on multiple machines using message passing interface (MPI) to communicate.
\item 
{\bf BIDMach}~\cite{bidmach2015}: BIDMach is a single GPU library that contains a set of matrix functions on top of which machine learning algorithms can be built. It also implements ALS based on a general purpose sparse matrix function.
\item 
{\bf HPC-ALS}~\cite{als2015}: It implements ALS on single GPU by exploiting registers and shared memory. However, it has no non-coalesced read, approximate solver or reduced precision.
\item 
{\bf GPU-ALS}~\cite{hpdc2016}: The state-of-art ALS implementation on GPUs but without our memory optimization and approximation presented in Section\ref{sec:memory} and Section\ref{sec:solver}, respectively. 
\item
{\bf GPU-SGD}~\cite{cumfsgd}: Section~\ref{sec:pre} discussed the difference between ALS and SGD solvers for MF.  We also compared with a CUDA-based SGD solution that solves MF problems with one or multiple GPUs, using matrix blocking and Hogwild!-style algorithms~\cite{hogwild-nips11} to parallelize the SGD updates. For individual SGD updates, it leverages GPU architectural features such as cache, warp-shuffle instructions, and half-precision floats. 
%comment the following sentence if we want to comment out als vs sdg                 
In Section~\ref{sec:exp-alssgd} we will discuss and compare these two methods in detail.
\end{itemize}

It is worth mentioning that, the performance of CPU-based algorithms not necessarily improves as the number of threads/machines increases due to two reasons: 1) synchronization on shared data structures and 2) communication overhead.  Enlarging the number of compute resource may even hurt their performance~\cite{als2015}. Therefore, we use 40 threads for LIBMF, which achieves the best performance. For NOMAD, we use the best settings as reported in~\cite{nomad14}, which are 32 machines for Netflix and Yahoo, and 64 machines for Hugewiki. Moreover, BIDMach and HPC-ALS can only use one GPU, while GPU-ALS and \algname can adopt multiple GPU settings. We test all GPU-based algorithms using one GPU on both Netflix and YahooMusic.  Furthermore, to show how well both GPU-ALS and our framework scale with the number of GPUs, we use four GPUs for both algorithms on the Hugewiki dataset.  

Figure \ref{fig:speed} shows the relation between test RMSE and training time while table \ref{table:time} summarizes the time when RMSE reaches an acceptable level. Clearly, \algname outperforms all CPU solutions with a large margin. Specifically, on both Netflix and YahooMusic datasets, \algname with single Pascal GPU (\algnamens@P) achieves 5.6x-7x performance gain compared to LIBMF. As for Hugewiki, \algname with four Pascal GPUs only takes 68 seconds to converge, which is significantly faster compared to 459 seconds for NOMAD (6.7x) and 3021 seconds for LIBMF (44.4x).  The reason BIDMach is not included in the table is that it does not converge to the acceptance level.  Regardless the convergence, we can observe the ALS kernel of BIDMach runs at 40 GFLOPS, which is similar to the reported measurement in their original paper \cite{bidmach2015}.  However, 40 GFLOPS is much lower than \algname (see section~\ref{sec:exp-hw} for our performance in terms of FLOPS).  On the other hand, since HPC-ALS is not open-source, we only compare our performance of per iteration time on Netflix, which has been reported in their paper.  Results show that \algname runs twice as fast as HPC-ALS on the same hardware (Kepler K40). Furthermore, compared with GPU-ALS, \algname has a significant performance advantage thanks to our memory optimization and approximate computing techniques. On Netflix with Maxwell GPU, \algname only needs 6.5 seconds to converge while GPU-ALS needs 28 seconds, \textit{i.e.} a 4x speedup. As a summary, \algname also outperforms all state-of-art GPU solutions.

\begin{table}[t]
    \begin{center}
      \caption{Training time in seconds when converging to acceptable RMSE. @M: Maxwell GPU, @P: Pascal GPU.}
    \begin{tabular}{| >{\centering}p{2.2cm} | >{\centering}p{1.5cm} |>{\centering}p{1.7cm}| >{\centering}p{1.5cm} | >{\centering}p{1.5cm} |}
    \hline
     \textnormal{Studies}     & \textit{Netflix} & \textit{YahooMusic} & \multicolumn{2}{c|}{\textit{Hugewiki}}  \\ \hline
     {LIBMF~\cite{libmf++}}    & 23 & 38 & \multicolumn{2}{c|}{3021} \\ \hline
     {NOMAD~\cite{nomad14}}    & 9.6 & 109 & \multicolumn{2}{c|}{459}  \\ \hline

%{MLGF-MF} & $>80$ &$>300$ & \multicolumn{2}{c|}{$\sim 1000$} \\ \hline
{GPU-ALS@M~\cite{hpdc2016}} & $28$   & $42$ & \multicolumn{2}{c|}{400} \\ \hline
{\algnamens@M} & 6.5  & 13.2 &\multicolumn{2}{c|}{166}\\ \hline
{\algnamens@P} & 3.3 & 6.8 &\multicolumn{2}{c|}{68} \\ \hline
{\algnamens@P\newline /LIBMF} & 7x & 5.6x &\multicolumn{2}{c|}{44.4x}            
\\ \hline
  \end{tabular}
  \label{table:time}
  \end{center}
\end{table}

%%%%%%%%%%%%%%%%%%%%%%%%%%%%%%%%%%%%%%%%%%%%%%%%%%%%
%%%%%%%%%%%%%%%%%%%%%%%%%%%%%%%%%%%%%%%%%%%%%%%%%%%%
\subsection{Has \algname fully exploited GPU?}
\label{sec:exp-hw}

In this section, we validate whether the proposed \algname framework has fully exploited the potential of GPU hardware.  The analytics are done by examining if the compute-intensive kernel \texttt{get\_hermitian} has achieved high FLOPS, and if the memory-intensive CG solver has achieved high memory bandwidth.

\vspace*{0.05in}
\noindent \textbf{Has \texttt{get\_hermitian} achieved high FLOPS?}
\vspace*{0.05in}

\noindent To obtain \texttt{get\_hermitian}: ${\small A_u=\sum\limits_{r_{uv}\neq0}(\boldsymbol\theta_v \boldsymbol\theta_v^T+\lambda I)}$ for $1 \leq u \leq m$, one needs to read the sparse matrix $R$ and perform $m$ matrix multiplications.  The of size of each multiplication is $\mathbb R^{f\times n_{x_u}}\times \mathbb R^{n_{x_u}\times f}$, where $n_{x_1}+n_{x_2}+...+n_{x_m}=N_z$.  To our best knowledge, no existing GPU library including cuBLAS has implemented the \texttt{get\_hermitian} function for us to compare.  The closest baseline is the batched-matrix-multiplication \texttt{gemmBatched}~\cite{cublas-gemmbatched} in cuBLAS, which calculates $m$ matrix multiplications of the same dimension: $\mathbb R^{a\times b}\times \mathbb R^{b\times c}$.  Although \texttt{gemmBatched} cannot parallelize matrix multiplications with different sizes, to compare, we set the dimension of each computation in our \texttt{get\_hermitian} to be the same. Under this setting, two algorithms can be fairly compared and we measure the FLOPS achieved by both with Kepler, Maxwell and Pascal on Netflix.

\begin{figure}[t]
\center{\includegraphics[width=\linewidth]
        {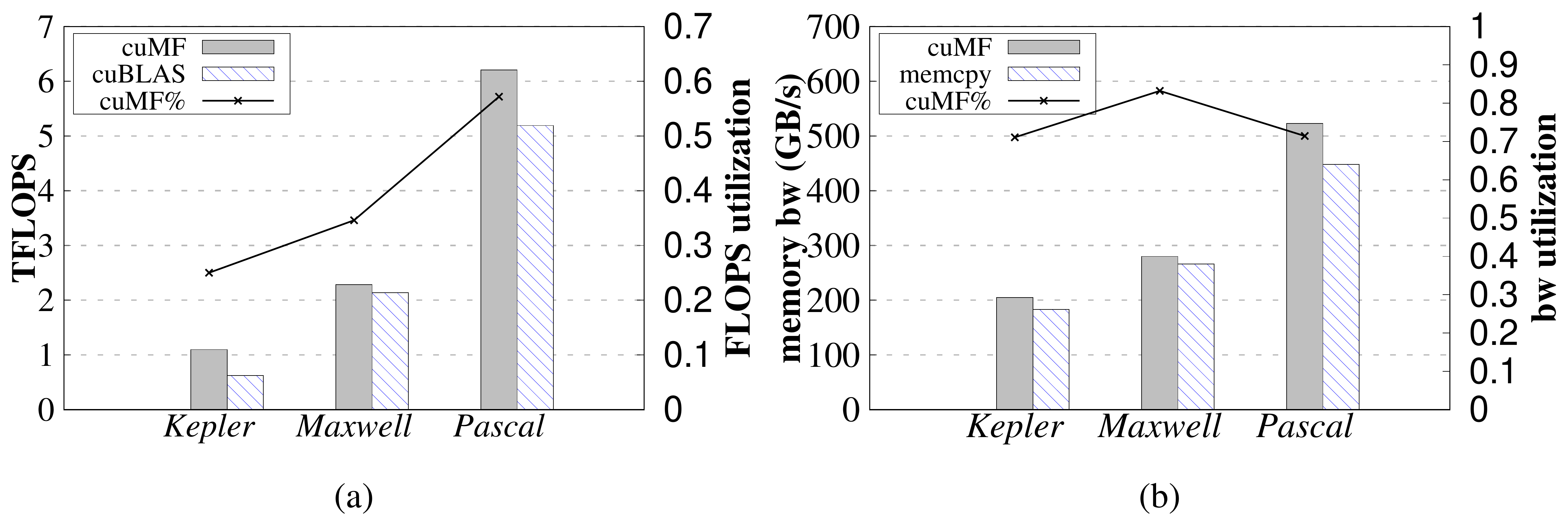}}
 \caption{\textbf{ (a) The FLOPS and efficiency of \texttt{get\_hermitian}} on GPUs of three generations. \algname achieves higher FLOPS than the batch cuBLAS routine for fixed size and higher FLOPS efficiency on newer GPUs. (b) The memory bandwidth achieved by CG solver is shown to be higher than the bandwidth of \texttt{cudaMemcpy}.}
\label{fig:flopsbw}
\end{figure}

Experimental results, as illustrated in Figure~\ref{fig:flopsbw}(a), shows that \algname achieves higher FLOPS in all three generations of GPUs.  This is impressive because \texttt{get\_hermitian} compared to \texttt{gemmBatched} in cuBLAS needs to perform extra work. Specifically, \texttt{get\_hermitian} needs to read sparse $R$ to get references to $\Theta$, and batch-multiply matrices with variable sizes. However, cuBLAS only needs to read dense input and batch-multiply matrices with the same size, which has no time cost on finding references.  Moreover, regarding FLOPS efficiency (\emph{i.e.}, the achieved-FLOPS divided by the device's peak-FLOPS) Figure \ref{fig:flopsbw}(a) indicates that \algname achieves better performance in Nvidia newly developed architectures.  This can be explained by the fact that performance of \texttt{get\_hermitian} is generally limited by the number of registers. Comparing Kepler, Maxwell and Pascal, the number of registers per core increases as technology evolves. At the same time, it reveals that our design discipline matches the development trend of GPU.

\vspace*{0.05in}
\noindent \textbf{Has the CG solver achieved high memory bandwidth?} 
\vspace*{0.05in}

\noindent We measure the memory transfer rate (GB/s) between GPU SMs and its DRAM, and compare it to the bandwidth achieved by CUDA function \texttt{cudaMemcpy}. Since \texttt{cudaMemcpy} only copies memory and deals with no computation, the comparison with it can indicate how well the CG solver can saturate the device memory bandwidth. As seen from Figure~\ref{fig:flopsbw}(b), \algname achieves a higher bandwidth than \texttt{cudaMemcpy} on all three types of GPUs. This demonstrates that our proposed CG solver utilizes the memory bandwidth efficiently.

%%%%%%%%%%%%%%%%%%%%%%%%%%%%%%%%%%%%%%%%%%%%%%%%%%%%%%%%%%%%%%%%%%%%%%%%%%%%%%
%CG inner iteration number tuning.
\subsection{ALS vs. SGD on GPUs}
\label{sec:exp-alssgd}

\begin{figure*}[h]
\centering
\includegraphics[scale=0.37,angle=0]{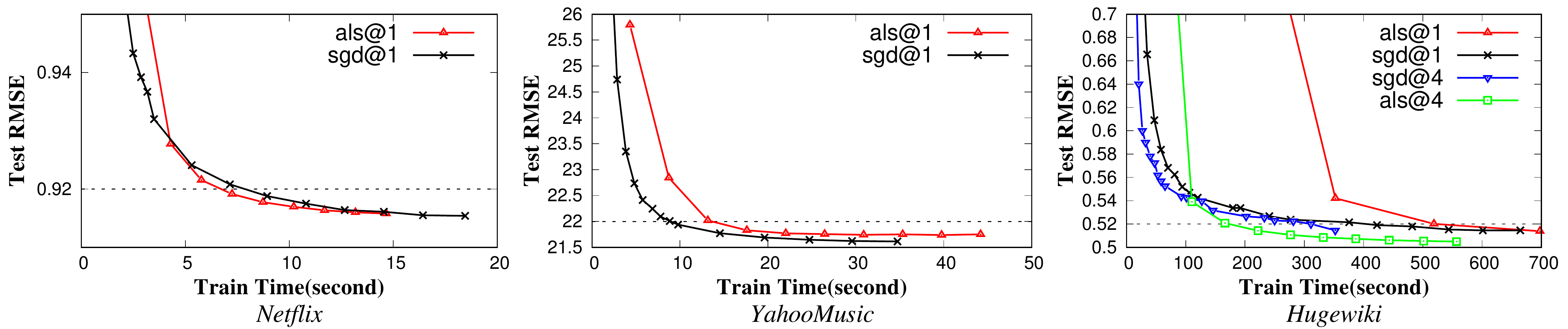}
\caption{\textbf{ALS vs. an SGD solution~\cite{cumfsgd} on one (@1) and four (@4) GPUs.}}\label{figure:sgdals}
\end{figure*}

As discussed in early Section \ref{sec:pre}, ALS and SGD have their own attributes in solving the problem of MF. SGD runs faster per iteration but requires more iterations. When the rating matrix gets denser, ALS has more advantage because SGD's complexity grows~\cite{mf-computer09} and also becomes harder to parallelize~\cite{cumfsgd}. 

We implemented both ALS and SGD\footnote{\url{https://github.com/cuMF/}} and compare their performance on GPUs.  We follow the same setting as before, and report the results in Figure \ref{figure:sgdals}.  Come as no surprise, ALS runs slower in each iteration, but requires fewer iterations to coverage. On one GPU, ALS converges slightly faster than SGD on Netflix, but slightly slower than SGD on YahooMusic and Hugewiki. However, with four GPUs (als@4), ALS converges faster than SGD on Hugewiki data.

Moreover, as seen in Table~\ref{tbl:complexity}, SGD's computation complexity ($\mathcal O(N_zf)$) grows linearly with $N_z$. This makes it inefficient when the rating matrix $R$ becomes more dense. This issue becomes severe when dealing with implicit inputs, where the rating matrix is considered fully dense, i.e., $N_z=m*n$~\cite{imf2008}. ALS can easily adapt to the setting of MF with implicit inputs, which we will discuss separately in section \ref{sec:imf}.  

% We compare with a single GPU solution that is based on SGD~\cite{cumf_sgd}. The test RMSE curves of \algname and the SGD solution on three datasets running on Maxwell are depicted in Figure~\ref{figure:sgdals}. Come as no surprise, ALS runs slower in each iteration, but requires fewer iterations to coverage. In terms of wall-clock time, ALS is slightly faster than SGD on Netflix, but slightly slower than SGD on YahooMusic and Hugewiki, using one GPU.  

% Despite the fact that \algname can be slower than SGD on one GPU, it is still advantageous for two reasons. First, as discussed in Section~\ref{sec:pre}, ALS is easy to parallelize for independent processing on individual features. While SGD is inherently sequential and parallizing it is more challenging. Existing parallel SGD solutions either require intensive communication and therefore inefficient, or avoid communication by asychronous update but negatively impact convergence. As a result, the SGD solution~\cite{cumf_sgd} can only efficiently use one GPU. On the contrary, we have developed data and model parallelism schemes for ALS that can scale to multiple GPUs~\cite{hpdc2016}. As seen in Figure~\ref{figure:sgdals}, \algname with four GPUs (\textit{als@4}) is the fastest on Hugewiki. The second reason, which will be explained with more details in section~\ref{sec:imf}, is ALS' adaptability in solving the implicit MF problem. 

%%%%%%%%%%%%%%%%%%%%%%%%%%%%%%%%%%%%%%%%%%%%%%%%%%%%%%%%%%%%%%%%%%%%%%%%%%%%%%
\subsection{Implicit matrix factorization}
\label{sec:imf}
MF with implicit inputs has been widely used in real-life applications~\cite{spotify,imf2008} where explicit ratings are replaced by implicit ones such as purchase or number of clicks.  To show ALS is able to handle implicit inputs, we follow the same setting as shown in \cite{imf2008}, by considering a binary matrix $P \in \mathbb R^{m\times n}$. $p_{uv}=1$ if the implicit observations $r_{uv}>0$, and $p_{uv}=0$ otherwise. The original paper also adds confidence measures $c_{uv}$ to predictions, which leads to the following cost function:
\begin{equation*}
\small
 \min_{X, \Theta} \sum\limits_{u,v} c_{uv}(p_{uv} - \boldsymbol{x}_u^T\boldsymbol{\theta}_v)^2,
\label{eq-imf}
\end{equation*}
where $c_{uv}=1+\alpha r_{uv}$ and $\alpha$ is a given scaling constant. In other words, any $r_{uv}=0$ is no longer treated as a missing rating, but as a zero-rating with low confidence $c_{uv}$. Under this assumption \cite{imf2008}, $P$ is not sparse and therefore SGD will be costly.  In such a way, SGD loses its competitiveness.  Therefore, we compare \algname with two open-source libraries for implicit MF: \textit{implicit}\footnote{\url{http://github.com/benfred/implicit}} and \textit{QMF}\footnote{\url{http://github.com/quora/qmf}}.  Experiments demonstrate that \algname converges under the implicit setting and the per iteration time of \algname, \textit{implicit} and \textit{QMF} are 2.2, 90, and 360 seconds, respectively. 

\section{Related Work}
\begin{table*} [h]
\centering
\caption{\textbf{Parallel MF solutions using SGD, ALS and CCD, on CPUs and GPUs.}}
\label{tbl:relatedwork}
\begin{tabular}{|m{0.05\textwidth}|p{0.52
\textwidth}|p{0.36\textwidth}|}  \hline
\textbf{}& \multicolumn{1}{c|}{\textbf{CPU}} & \multicolumn{1}{c|}{\textbf{GPU}} \\ \hline
\textbf{SGD} & \textbf{lock-free}: workers independently sample \& update \newline
\phantom{x}single-node: HogWild!~\cite{hogwild-nips11}; multi-nodes: FactorBird~\cite{factorbird14}, Petuum~\cite{Xing2014-PS}  \newline 
\textbf{blocking}: workers pick non-overlapping blocks \newline
\phantom{x}blockDim=\#workers: DSGD~\cite{DSGD-kdd11} \newline
\phantom{x}blockDim$>$\#workers: LIBMF~\cite{libmf-13}, NOMAD~\cite{nomad14}, DSGD++~\cite{DBLP:conf/icdm/TeflioudiMG12} \newline
\phantom{x}nested blocking: dcMF~\cite{dcMF15}, MLGF-MF~\cite{kdd15mf}
& \textbf{single and multiple GPUs}: GPU-SGD -- SGD with lock-free and blocking~\cite{cumfsgd} \\ \hline
\textbf{ALS} & \textbf{replicate all features}: PALS~\cite{netflix08},  DALS~\cite{DBLP:conf/icdm/TeflioudiMG12} \newline
\textbf{partial replicate features}: SparkALS~\cite{sparkmllib16}, GraphLab~\cite{graphlab12},  Sparkler~\cite{sparkler13} \newline
\textbf{rotate features}: Facebook~\cite{facebook15} \newline
\textbf{approximate ALS}: \cite{als-10}
& \textbf{single GPU}: BIDMach~\cite{bidmach2015}, HPC-ALS~\cite{als2015} \newline
\textbf{single and multiple GPUs}: GPU-ALS~\cite{hpdc2016} and \algname
\\ \hline
\textbf{CCD} & \textbf{multi-core and multi node}: CCD++~\cite{ccd++-icdm12} & \textbf{single GPU}: parallel CCD++~\cite{ccd++17}\\ \hline
\end{tabular}
\end{table*}
This section reviews related work on parallel matrix factorization with SGD, ALS and cyclic coordinate descent (CCD) algorithms. Table~\ref{tbl:relatedwork} is a summary and details are in the following subsections.
%%%%%%%%%%%%%%%%%%%%%%%%%%%%%%%%%%%%%%%%%%%%%%%%%%%%%%%%%%%%%%%%%
%%%%%%%%%%%%%%%%%%%%%%%%%%%%%%%%%%%%%%%%%%%%%%%%%%%%%%%%%%%%%%%%%
\subsection{Parallel SGD}
SGD is inherently serial where each time one sample is selected to update. To accelerate this process, two samples can update in parallel if they are neither in same row nor same column. This observation has led to two ways to parallel SGD for MF: lock-free Hogwild!~\cite{hogwild-nips11} and blocking~\cite{ccd++-icdm12,libmf-13,DSGD-kdd11,kdd15mf}. \textit{Hogwild!} observes that when $R$ is very sparse and the number of parallel workers is much less than the dimension of $R$, they can independently update samples with a low probability of conflict. \textit{Blocking} divides $R$ into several sub-blocks, and sub-blocks that do not share rows or columns can update in parallel.

\vspace*{0.05in}
\noindent\textbf{CPU approaches.} SGD has been parallelized in multi-core \cite{libmf-13,kdd15mf}, multi-node MPI \cite{DBLP:conf/icdm/TeflioudiMG12, nomad14}, MapReduce \cite{DSGD-kdd11} and parameter-server \cite{factorbird14, Xing2014-PS} systems. These methods partition $R$ into blocks with no overlapping rows or columns, and work on these blocks in parallel. They further optimize the algorithm with asynchronous communication, overlapping  communication and computation, and shared memory. For example, LIBMF \cite{libmf-13} is very efficient on multi-cores.
%It has out performed nearly all other approaches on a 12-core machine.
However, it stops scaling when using few dozens cores~\cite{cumfsgd,dcMF15}, because of the locking in a shared data structure. Moreover, LIBMF is a single-machine solution and therefore cannot deal with large-scale problems. NOMAD \cite{nomad14} extends the idea of block partitioning, and alleviate the issue of global locking. It performs similarly to LIBMF on a single machine and can scale to a 64-node HPC cluster.
%\textbf{Parameter Server with SGD.} 
%More recently, the idea of ``parameter server" \cite{Smola2014-PS, Xing2014-PS} emerges for extremely large-scale machine learning problems. In this paradigm, the \textit{server nodes} store parameters, while the \textit{worker nodes} store training data and compute on them. The parameter-server framework manages asynchronous communication between nodes, flexible consistency models, elastic scalability, and fault tolerance. Following this idea, 
Parameter server~\cite{Xing2014-PS} can be used to implement distributed SGD. For example, Petuum \cite{Xing2014-PS} can scale MF to hundreds of cores in a cluster, and Factorbird \cite{factorbird14} is a parameter server specifically implemented for matrix factorization. 

\vspace*{0.05in}
\noindent\textbf{GPU approaches.} Both Hogwild and blocking schemes are implemented in~\cite{cumfsgd}. It has efficient kernels for SGD update, leveraging cache, warp-shuffle instructions, and half-precision. 
%cuMF\_sgd outperforms all other approaches using one GPU card.

%%%%%%%%%%%%%%%%%%%%%%%%%%%%%%%%%%%%%%%%%%%%%%%%%%%%%%%%%%%%%%%%%
%%%%%%%%%%%%%%%%%%%%%%%%%%%%%%%%%%%%%%%%%%%%%%%%%%%%%%%%%%%%%%%%%

\subsection{Parallel ALS and CCD} 
\noindent\textbf{CPU approaches for ALS.} PALS \cite{netflix08} and SparkALS \cite{sparkmllib16} parallelize ALS by feature full replication and partial replication, respectively. These approaches are not feasible when feature matrices get extremely large. Facebook \cite{facebook15} tackles this issue by partitioning the feature matrix and rotate its parts among multiple nodes. 
%For example, when solving $X$, $X$ is partitioned disjointedly across nodes; $\Theta$ is also partitioned and rotated across the same set of nodes. 
%When a $\Theta$ partition $\Theta^{(j)}$ meets $X$ partition $X^{(i)}$, $X^{(i)}$ is updated by observing $\Theta^{(j)}$; $X^{(i)}$ completes an iteration of update after it meets all $\Theta^{(j)}$s. 
%This is somewhat similar to SU-ALS but SU-ALS does not use rotation, as GPUs do not have sufficient memory to do rotation.
GraphLab \cite{graphlab12} distributes the feature matrix among multiple machines. When updating in a machine, needed features are fetched on-demand from other machines. %This involves a lot of cross-node traffic and puts a high requirement on network bandwidth.

% \cite{Zastrau:2012:SGD} implements both SGD and ALS on GPU to solve MF. It uses a mini-batch-based and sequential version of SGD, and a variant of ALS that adjusts (rather than re-calculates) the inverse of the Hermitian matrices in each iteration. They neither optimize the memory access to fully utilize GPU's compute power, nor scale to multiple GPUs to handle large-scale problems. 
\vspace*{0.05in}
\noindent\textbf{GPU approaches for ALS.} BIDMach~\cite{bidmach2015} provides generic matrix kernels for many machine learning algorithms including MF. However, its sparse kernel is not specifically optimized for ALS and slower than \algname. HPC-ALS~\cite{als2015} optimizes the \texttt{get\_hermitian} kernel similar to us. However, they used neither non-coalesced read, nor approximate solver nor reduced precision.
%%%%%%%%%%%%%%%%%%%%%%%%%%%%%%%%%%%%%%%%%%%%%%%%%%%%%%%%%%%%%%%%%
%%%%%%%%%%%%%%%%%%%%%%%%%%%%%%%%%%%%%%%%%%%%%%%%%%%%%%%%%%%%%%%%%

\vspace*{0.05in}
\noindent\textbf{Parallel CCD}. CCD++ \cite{ccd++-icdm12} performs sequential updates on one row of the decomposed matrix while fixing other variables. CCD++ has lower time complexity but makes less progress per iteration, compared with ALS. ~\cite{ccd++17} further accelerates CCD++ on GPUs using loop fusion and tiling. The resulting algorithm is shown to be faster than CCD++ on CPUs \cite{ccd++-icdm12} as well as GPU-ALS~\cite{hpdc2016} that is without memory optimization and approximate computing.

%In practice, CCD++ behaves well in the early stage of optimization, but then becomes slower than libMF.

%%%%%%%%%%%%%%%%%%%%%%%%%%%%%%%%%%%%%%%%%%%%
\section{Conclusion}

Due to the importance of MF in the field of data mining, in this paper, we accelerate ALS, one of the most important MF solving algorithms with GPUs.  Specifically, we identify challenges that make ALS running slow such as constraints in the utilization of capacity and bandwidth of memory, and computation intensiveness.  To alleviate these problems, we propose a novel framework named \algnamens.  Our algorithm exploits the GPU memory architectures and shortens the time of reading data via an innovative scheme.  At the same time, the proposed algorithm also eliminates unnecessary computing in solving MF without hurting convergence.  We conduct extensive experiments under various settings. Our proposed method achieves the state-of-the-art performance, suggesting that \algname can significantly advance the task of MF. We also integrated \algnamens\ into Spark MLlib, accelerating its ALS algorithm\footnote{\url{https://github.com/IBMSparkGPU/CUDA-MLlib}}. 

In future work, we would like to further analyze different GPU-accelerated MF algorithms and investigate algorithm selection based on dataset characteristics such as dimensions and sparsity, and hardware resource constraints such as number of GPUs. We also plan to investigate a hybrid solution that combines SGD and ALS. A scenario could be using ALS for the initial batch training and SGD for incremental updates of the model.  Last but not the least, we would like to exploit the new Nvidia Tensor Cores~\cite{tensorcores} hardware that natively supports half-precision arithmetic, to further speed up \algname.

% Matrix factorization is a key to many algorithms including collaborative filtering, word embedding and compression. GPUs enable us to consolidate big compute FLOPS and memory bandwidth on one or few machines, which makes it desirable to solve matrix factorization. We accelerate the two dominant functions in ALS-based matrix factorization: \texttt{get\_hermitian} and \texttt{solve}. 
% We implement cuMF\_als in a shared-memory system with multiple GPUs. Experiments show that, cuMF\_als outperforms all competing CPU solutions by a large margin. Its performance is improved by \textbf{2x-4x} compared with a state-of-the-art GPU solution, thanks to the memory optimization and approximate computing techniques. We believe that our main optimization strategy can be useful for other machine learning algorithms, especially those on bipartite graphs.

\bibliographystyle{abbrv}
\bibliography{main}

\end{document}